\documentclass[a4paper,aps,prl,amsmath,amssymb,floatfix,twocolumn,reprint,showpacs,superscriptaddress]{revtex4-1}
\usepackage{graphicx}
\usepackage{dcolumn}
\usepackage{bm}

\begin{document}

\title{Understanding metropolitan patterns of daily encounters}

\author{Lijun Sun}
\affiliation{Future Cities Laboratory, Singapore-ETH Centre for Global Environmental Sustainability (SEC), 138602, Singapore}
\affiliation{Department of Civil \& Environmental Engineering, National University of Singapore, 117576, Singapore}
 \author{Kay W. Axhausen}
\affiliation{Future Cities Laboratory, Singapore-ETH Centre for Global Environmental Sustainability (SEC), 138602, Singapore}
\affiliation{Institute for Transport Planning and Systems (IVT), Swiss Federal Institute of Technology, Z\"{u}rich, 8903, Switzerland}

\author{Der-Horng Lee}
\affiliation{Department of Civil \& Environmental Engineering, National University of Singapore, 117576, Singapore}

\author{Xianfeng Huang}
\affiliation{Future Cities Laboratory, Singapore-ETH Centre for Global Environmental Sustainability (SEC), 138602, Singapore}
\affiliation{LIESMARS, Wuhan University, Wuhan, 430079, China }

\date{\today}

\begin{abstract}
Understanding of the mechanisms driving our daily face-to-face encounters is still limited; the field lacks large-scale datasets describing both individual behaviors and their collective interactions. However, here, with the help of travel smart card data, we uncover such encounter mechanisms and structures by constructing a time-resolved in-vehicle social encounter network on public buses in a city (about 5 million residents). This is the first time that such a large network of encounters has been identified and analyzed. Using a population scale dataset, we find physical encounters display reproducible temporal patterns, indicating that repeated encounters are regular and identical. On an individual scale, we find that collective regularities dominate distinct encounters' bounded nature. An individual's encounter capability is rooted in his/her daily behavioral regularity, explaining the emergence of ``familiar strangers'' in daily life. Strikingly, we find individuals with repeated encounters are not grouped into small communities, but become strongly connected over time, resulting in a large, but imperceptible, small-world contact network or ``structure of co-presence'' across the whole metropolitan area. Revealing the encounter pattern and identifying this large-scale contact network are crucial to understanding the dynamics in patterns of social acquaintances, collective human behaviors, and -- particularly -- disclosing the impact of human behavior on various diffusion/spreading processes.

\end{abstract}

\maketitle

\section{Introduction}

Highlighting their importance in various temporal spreading processes \cite{Holme2012,Perra2012,Krings2012}, recent studies of human contact networks demonstrate an increasing interest in physical encounters \cite{Read2008,Stehle2011,Salathe2010,Isella2011,Rocha2011,Grannis2009}. Contrary to non-physical social contacts initiated by mobile phone calls, emails and online social networks \cite{Eckmann2004,Barabasi2005,Onnela2007,Golder2011,Rybski2012}, human subjects' physical encounters take place with heterogeneous prior knowledge of each other, from acquaintances to unknowns, linking two individuals by their co-presence in both spatial and temporal dimensions \cite{Crandall2010}. On the other hand, with increasing human interactions, communities may also emerge from social contagion enabled by physical proximity: from not noticing each other, to unintentionally interacting, to intentional communicating, to mutual trust \cite{Grannis2009}. We are tracing -- for the first time and for a large population -- in this case for all of Singapore's bus users, how these encounters are structured by both individual behavior and institutional structures. Bus use is a small slice of urban life, but one where ``familiar strangers'' will emerge \cite{Milgram1974,Paulos2004} -- strangers who have been encountered frequently in daily life, but might never have been addressed. This context is one of many: which, in their totality, give residents the social background against which they construct their more intense relationships. We believe the joint encounter pattern is influenced by individual regularity \cite{Gonzalez2008,Song2010}. A previous study based on the dispersal of bank notes suggested that human trajectories follow continuous-time random walks \cite{Brockmann2006}; however, considering the inherent regularity in individual behaviors, recent analyses of large-scale trajectories from mobile phone data and travel diaries indicate, on the contrary, that individual mobility patterns display significant regularity and remarkable predictability \cite{Gonzalez2008,Song2010,Schofelder2010}.

With the help of sensors and online networks, data describing close proximity in real-world situations sheds new light on encounter patterns and spreading dynamics in contact networks other than diary-based surveys \cite{Read2008}. However, these data collection systems are generally embedded in limited samples in spatially small-scale settings such as schools \cite{Salathe2010}, conferences and exhibitions \cite{Stehle2011,Isella2011}, and even in prostitution \cite{Rocha2011}. On a large scale, we still lack empirical data describing examples of both individual regularity and joint encounter patterns (other than simulating mobility and behavior patterns individually, relying on computational and agent-based models \cite{Eubank2004,Wang2009,Smieszek2011}). Thus, given data limitations, studies on individual mobility regularity and collective interactions are traditionally conducted separately: the mechanisms driving our daily encounters remain unclear.

\begin{figure*}[ht!]
\begin{center}
\includegraphics[width=.65\textwidth]{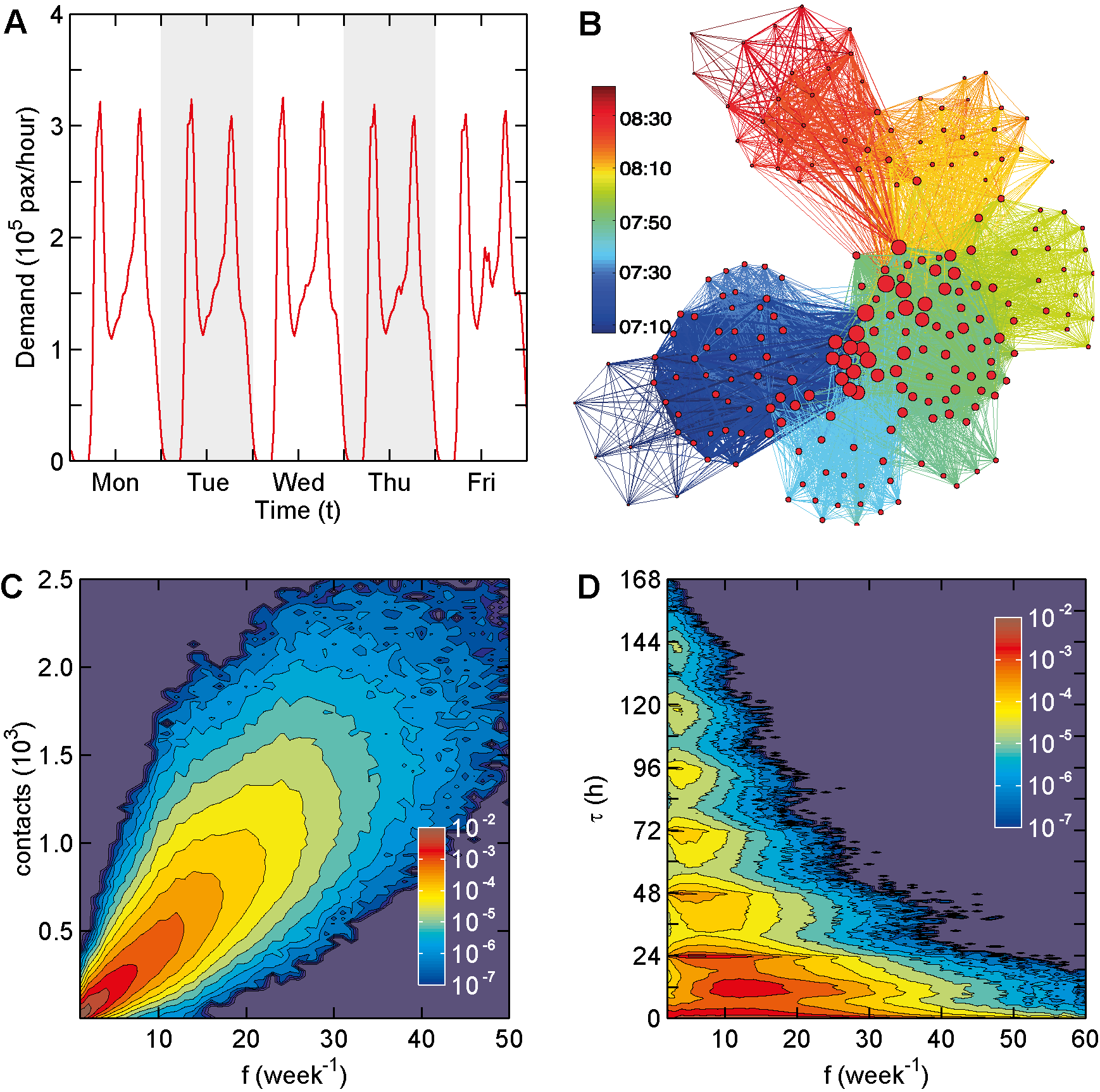}
\caption{Characteristics of transit behaviors. (A) Departure rate of city bus trips. Demand shows similar shapes from Monday to Friday. Two prominent peaks can be identified every day, indicating collective commuting behavior. (B) Time-resolved encounter network on one vehicle service (214 passengers) created using transaction records. Size of each node indicates number of encountered people; color and thickness of each edge indicate start time and duration - respectively - of each encounter activity. This figure explains usage of smart card data to create an empirical encounter network in city-scale. (C) Probability density function $P\left(f,n\right)$ of trip frequency $f$ and number of encountered people $n$ of all individuals. Symmetric pear-shaped pattern suggests linear relationship of $n = \sum\nolimits_{i = 1}^f {{n_i} \approx 50f}$. (D) Probability density function $P\left(f,\tau\right)$ of trip frequency $f$ and inter-event interval $\tau$ between consecutive bus trips. Although diversity of individual transit usage obscures the distribution, a remarkable area with $\left( {5 \le f \le 25,0 \le \tau  \le 25} \right)$ can be identified through similarity in regular commuters' transit behavior.}
\label{Fig1}
\end{center}
\end{figure*}

Therefore, with the increasing quantity and range of human mobility, a central task is to explore social interaction patterns along with mobility regularity. However, previous data collection techniques fail to offer high-resolution information on collective interactions on a large scale (across the population). In this context, individual-based passive data collections embedded in our daily life, such as credit cards and smart cards transactions, can be advantageous. At present, transit usage might be the best proxy to capture the patterns of both individual mobility and collective interactions in an urban environment \cite{Pelletier2011}. Here, we use public transit transaction records to uncover encounter patterns (see SI Appendix section I for a detailed description). This dataset consists of more than 20 million bus trips from 2,895,750 anonymous users over one week (Fig.~\ref{Fig1}A) (about 55\% of the resident population) in Singapore. The high spatial-temporal resolution of this dataset allows us to extract time-resolved in-vehicle encounters, defined as two individuals occupying the same vehicle simultaneously (Fig.~\ref{Fig1}B). Using these, a city-scale empirical temporal contact network is created.

\section{Results}

Use of transit service in general, and buses in particular, is differentiated by ethnicity, gender, age and income, meaning that daily transit use might exhibit social segregation as well. To address dependency and segregation of bus use on social demographic attributes, we incorporated two additional datasets aside from smart card transactions: population census and national household interview travel survey (HITS). For our case, although transit use in Singapore shows dependency on age and income, public transit is still the most important transport means for daily commuting trips across all Singaporeans (see SI Appendix section II). By studying transit usage, we found that both trip duration and trip frequency can be accurately characterized by exponentially decaying tails, showing that people's transit activities are limited in number and duration during one week (see SI Appendix section III.2 and IV.1). To understand individuals' transit use patterns, we first measured the number of encountered people $n$ against trip frequencies $f$ for each individual. We find that the joint distribution $P\left(f,n\right)$ has a symmetric pattern against $n/f\approx 50$, indicating the substantial number of people encountered in each trip (Fig.~\ref{Fig1}C). We then measured the inter-event time $\tau$ between consecutive bus trips for the population, finding that $P\left(\tau\right)$ shows clear temporal patterns with prominent peaks (see SI Appendix section IV.3), which is contrary to the non-Poissonian nature of $P\left(\tau\right)\sim \tau^{-\beta}\exp\left(-\tau/\tau_c\right)$ in communication activities observed from digital networks \cite{Eckmann2004,Barabasi2005,Onnela2007,Golder2011,Rybski2012}, suggesting the periodicity of transit usage on the population scale. To explore the pattern of $\tau$ at the individual level, we grouped people according to their trip frequencies. In Fig.~\ref{Fig1}D, we added $f$ to each $\tau$ as an attribute and measured the joint probability distribution $P\left(f,\tau\right)$. Observing the heterogeneity rooted in individual behaviors helps us distinguish regular travelers from other passengers. Therefore, given the transit usage variation (and the analysis in SI Appendix section III and IV), we suggest that transit usage is a good example for capturing both individual mobility patterns and collective encounter patterns, certainly for our case study.

\begin{figure*}[!ht]
\begin{center}
\centerline{\includegraphics[width=.65\textwidth]{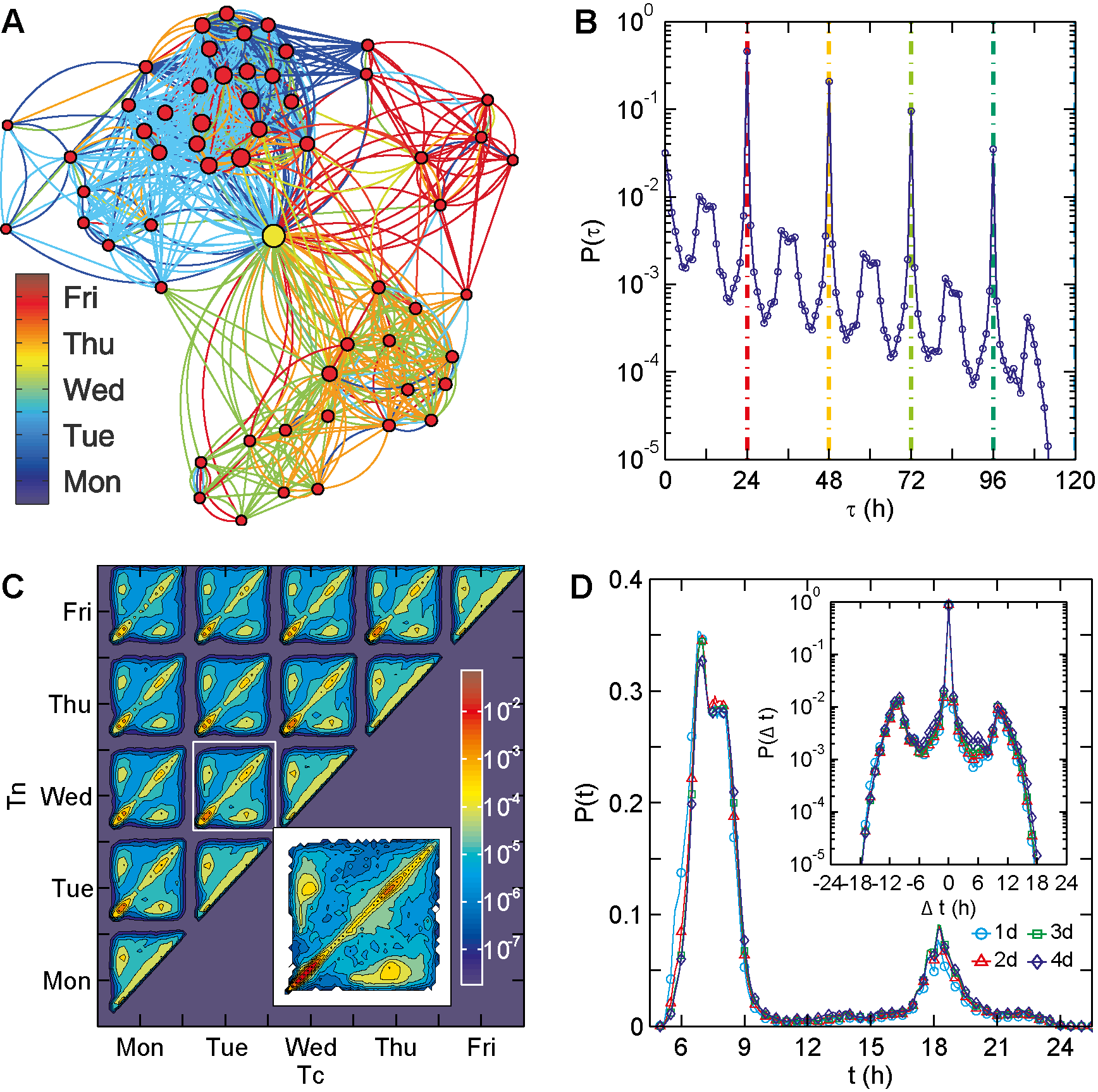}}
\caption{Temporal patterns of repeated encounters. (A) A typical temporal contact network of one individual with his/her ``familiar strangers'' (encountered more than once). Edge colors indicate start timings during the week. As observed, neighbors are also highly connected over time, suggesting network's strong social connection and small-world property. (B) Probability density function $P\left(\tau_e\right)$ of inter-event time between two consecutive encounters of paired individuals. The prominent peaks suggest that paired individuals tend to meet each other daily. A general decreasing trend is observed as well. On average, pairs of individuals with repeated encounters can meet each other about $27892055/18724388+1\approx 2.5$ times during the week. (C) Joint probability density $P\left(T_c,T_n\right)$ of timings of two consecutive encounters. Inset shows zoom-in of one density square. (D) Distribution $P\left(t\right)$ of encounter time for 4 groups with $D_n-D_c=\{1d,2d,3d,4d\}$, respectively. Inset of figures shows the distribution $P\left(\Delta t\right)$ of intervals between rescaled timing of paired encounters $\Delta t = t_n-t_c$. The prominent peak at $\Delta t=0$ suggests most repeated encounters happen at the same time.}
\label{Fig2}
\end{center}
\end{figure*}

As mentioned, an important phenomenon triggered by collective regularities is the ``familiar stranger'', which is also crucial in explaining how likely it is that the same persons will be encountered again (Fig.~\ref{Fig2}A). To explore the pattern of repeated encounters on the population scale, we created an aggregated encounter network over weekdays and measured the inter-event time $\tau_e$ between consecutive encounters of paired individuals, capturing 27,892,055 intervals from 18,724,388 pairs. We found that the distribution $P\left(\tau_e\right)$ is characterized by prominent peaks at $24h$, $48h$, $72h$ and $96h$, displaying a strong tendency of periodic encounters covering about 75\% of all cases, suggesting that the joint regularity also displays significant temporal periodicity (Fig.~\ref{Fig2}B). We also observed a decreasing pattern of $P\left(\tau_e\right)$, since the observed numbers decrease when interval $\tau_e$ is longer. To avoid bias, we extracted and grouped the timings of current and the next encounters $\left(T_c,T_n\right)$ over all pairs, so that for one pair with 3 encounters, 2 records are created ($\left(T_{1st},T_{2nd}\right)$ and $\left(T_{2nd},T_{3rd}\right)$). We then measured the joint distribution $P\left(T_c,T_n\right)$ (Fig.~\ref{Fig2}C). Strikingly, we found that the joint distribution presents reproducible patterns on the same diagonal with $D_n-D_c=\{1d,2d,3d\}$ respectively, where $D_c$ and $D_n$ represent the days of $T_c$ and $T_n$ respectively, suggesting the homogeneity of daily encounters. To measure reproducibility, we summed the probability for each day-of-the-week pair, excluding the diagonal cases with $D_c=D_n$, obtaining the density matrix:
\begin{equation}\nonumber
P\left( {{D_c},{D_n}} \right) = \left[ {\begin{array}{*{20}{c}}
0&{0.141}&{0.086}&{0.061}&{0.046}\\
0&0&{0.148}&{0.091}&{0.062}\\
0&0&0&{0.145}&{0.085}\\
0&0&0&0&{0.135}\\
0&0&0&0&0
\end{array}} \right]
\end{equation}

Considering the values in $P\left(D_c,D_n\right)$, we found that repeated encounters over the population can be modeled well as a Bernoulli process with probability of success $P_{encounter}\approx 0.33$, which is another factor behind the decrease of $P\left(\tau_e\right)$.

To reveal the homogeneity of day-to-day encounters, we further studied $P\left(T_c,T_n\right)$. As the inset of Fig.~\ref{Fig2}C shows, we found a strong diagonal on $23h\le T_n-T_c \le 25h$, which covers 85\% of the cases, suggesting that most recurring encounters happen at about the same time of day. Despite this, we also observed two areas representing cross-period encounters, such that the first in the afternoon and the following in the morning on the next day. To compare the distributions of different day-of-the-week pairs $\left(D_c,D_n\right)$, we grouped the pairs according to $D_n-D_c=\{1d,2d,3d,4d\}$ respectively, and rescaled both $T_c$ and $T_n$ to time of day $t_c$ and $t_n$. Then, we measured the probability density $P\left(t\right)$ of encounter time by merging $t_c$ and $t_n$ (Fig.~\ref{Fig2}D) and the distribution of inter-event intervals $\Delta t = t_n-t_c$ (the inset of Fig.~\ref{Fig2}D). Taken together, we found both $P\left(t\right)$ and $P\left(\Delta t\right)$ for different groups share indistinguishable shapes, indicating that the daily encounters can be characterized by a general temporal pattern. Furthermore, although afternoon peaks are longer and higher than morning peaks in daily transit use (Fig.~\ref{Fig1}A and see SI Appendix Fig. S6b), we confirmed that repeated encounters tend to happen more often in the morning, suggesting that collective regularity is more pronounced in the morning than in the afternoon \cite{Schofelder2010}. In this contact network, it is also implied that spreading via repeated interactions is more likely in the morning than afternoon. In addition, the prominent peaks at $\Delta t =0$ indicate that the most probable time for a recurring encounter is the same as the previous one.

\begin{figure*}[!htbp]
\begin{center}
\includegraphics[width=.65\textwidth]{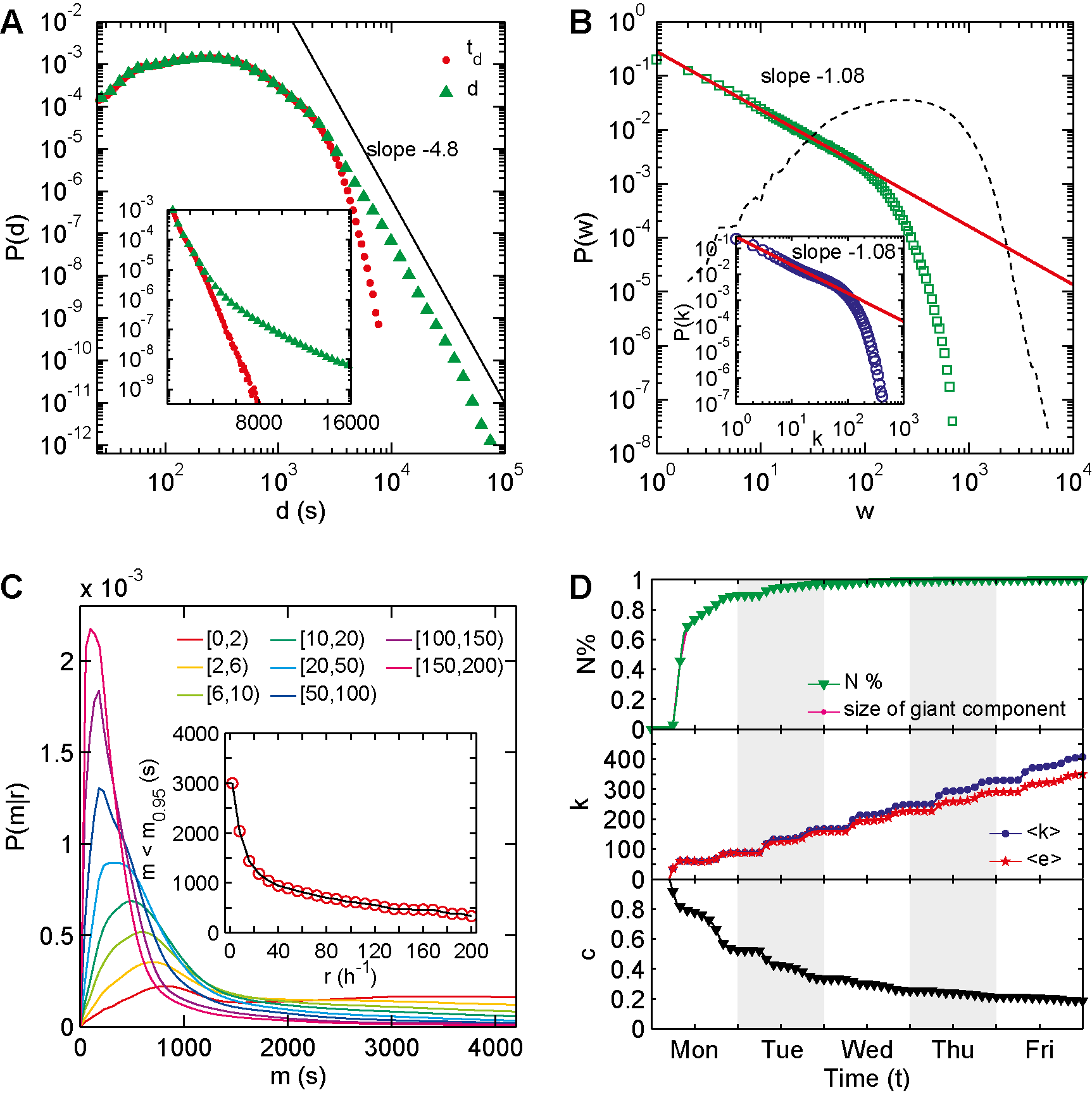}
\caption{Individual's impact on collective encounters. (A) Probability density functions $P\left(d\right)$ of sum of durations of repeated encounters and $P\left(t_d\right)$ of duration of individual encounters on weekdays; the same distributions are shown in semi-log scale as the inset. Solid line shows power-law with exponent $\beta=4.8$, suggesting that although distinct encounters exhibit exponentially bounded nature, paired total encounter duration presents scale-free property. Maximum values of $d$ and $t_d$ are $9.22\times 10^4s$ (25.6h) and $7.93\times 10^3s$ (2.2h), respectively. (B) Distributions $P\left(w\right)$ of personal weight and $P\left(k\right)$ of number of ``familiar strangers'' (inset), of 1,626,040 individuals with $w_i>0$. Solid lines show power-laws and dashed curve shows distribution $P\left(u\right)$ of number of total encounters ${u_i} \equiv \sum\nolimits_{j \in N\left( i \right)} {{f_e}\left( {i,j} \right)}$ as guides. (C) Probability density function $P\left(m\right)$ of absolute trip difference $m$ for grouped individuals with different $r$ (colored lines). Inset shows mean value of $m<m_{0.95}$ for people with certain $r$, suggesting a direct correlation: the higher one's encounter capability, the more stable one's behavior. (D) Evolution of time-aggregated network of 783,247 individuals with personal weight $w_i\ge 10$. Top inset shows the fraction of presented nodes and size of largest component. We found these two curves indistinguishable, suggesting high network connectivity. Middle inset shows evolutions of average number of encounters $\langle k \rangle $ and average number of encountered people $\langle e \rangle $. Bottom inset plots evolution of average clustering coefficient $c$ during the week, suggesting small-world property of this aggregated network.}
\label{Fig3}
\end{center}
\end{figure*}

Still, until now, the field lacked analyses to uncover mechanisms that drive encounters on an individual level. Therefore, we began to measure the encounter frequency of paired individuals, finding that the distribution $P\left(f_e\right)$ of encounter frequencies can be well characterized by a heavy-tailed distribution, even though the network is very dense, explaining the emergence of ``familiar strangers'' (see SI Appendix section VI.3). Although $f_e$ is a good approximation of connection strength, it fails to capture the actual overlapping of collective behaviors when considering other external factors such as transfers. To avoid misinterpreting the data, we used total encounter duration $d\left(i,j\right)=\sum\nolimits_{k = 1}^{{f_e}\left( {i,j} \right)} {{t_{d,k}}\left( {i,j} \right)}$ to better quantify the connection strength of individual pairs $\left(i,j\right)$, where ${t_{d,k}}\left( {i,j} \right)$ is the duration of their $k\,th$ encounter. In Fig.~\ref{Fig3}A, we show the distributions $P\left(d\right)$ (for all individual pairs) and $P\left(t_d\right)$ (over all trips) in both log-log scale and semi-log scale. We found that $P\left(t_d\right)$ can be well captured by an exponentially decaying tail, whereas $P\left(d\right)$ displays a power-law tail $P\left( d \right)\sim {d^{-\beta }}$ with exponent $\beta  \approx 4.8 \pm 0.1$. The significant degree of heterogeneity indicates that collective $d$ has overtaken the exponential bounded $t_d$. To summarize the observation of pairs of travelers, collective regularities do occur, suggesting that encounter patterns are influenced by paired individuals' behavior patterns.

Next, on an individual level (see SI Appendix section VII for modeling details), we propose a personal weight $w_i$ of individual $i$ proportional to their paired encounter frequency:
\begin{equation}
{w_i} \equiv \sum\nolimits_{j \in N\left( i \right)} {\left( {{f_e}\left( {i,j} \right) - 1} \right)}
\end{equation}
where $N\left( i \right)$ is the set of encountered people and ${f_e}\left( {i,j} \right)$ is the frequency of encounters between individual $i$ and $j$ observed. Thus, $w_i$ captures the likelihood of encountering ``familiar strangers''. In Fig.~\ref{Fig3}B, we chose individuals who had recurring encounters (${w_i} \ge 0$), and plotted the probability density functions $P\left( w \right)$ of personal weight and $P\left( k \right)$ of the number of ``familiar strangers'' respectively. We find that both distributions can be approximated well by power-laws with high cut-offs, with the same exponent $\beta  \approx 1.08 \pm 0.02$. It is important to note the great variation in person weights, i.e. encounter likelihoods, suggesting that encounter patterns might be influenced by individual behavior patterns.

To explore how individual behavioral regularity impacts collective encounter patterns, we tried to quantify both individual encounter capability and transit usage. First, to better measure individual encounter likelihood beyond travel time influence, we rescaled $w_i$ to ${r_i} = w_i/T_i \left(hour^{-1}\right)$ for each individual, where $T_i$ is the total travel time in hours. Inspired by the k-means clustering method -- for individual $i$ -- we used the absolute difference $m_i$ of morning and evening trips to quantify behavioral regularity:
\begin{equation}
  {m_i} = \sum\nolimits_{k = 1}^2 {\sum\nolimits_{{t_j} \in {S_k}} {\left| {{t_j} - {\mu _k}} \right| /n}}
\end{equation}
where $t_j=\left(t_{start}+t_{end}\right)/2$ is the mean of start and end times of the $j\,th$ ($j=1,2,\cdots,n$) trip and $\mu_k$ is the mean of $\left\{ {{t_j}|{t_j} \in {S_k}} \right\}$. Therefore, $m_i$ captures the time variation of daily transit use (the lower $m_i$ is, the more repetitive the individual will be; see SI Appendix section VII.2). To reveal the relation between behavioral regularity and encounter likelihood, we grouped the individuals according to $r_i$ and measured the distribution $P\left(m|r\right)$ for each group. As Fig.~\ref{Fig3}C shows, individuals with higher $r_i$ tend to have less skewed $P\left(m\right)$, whereas those with low $r_i$ display a more skewed distribution. In addition, we took the average of $m \le {m_{0.95}}$ over people with certain $r$, where $m_{0.95}$ is the $95^{th}$ percentile of their $m$, finding that the average absolute difference reaches about 50 min for $r < 4{h^{ - 1}}$, whereas for those with $r \ge 80{h^{ - 1}}$, the value decreases to less than 15 min, significantly shorter considering that the expected headway (i.e. service interval) of public buses is around 10 min (inset of Fig.~\ref{Fig3}C). In summary, we found that a larger encounter likelihood of an individual is strongly rooted in his/her behavioral regularity.

With these common daily physical interactions, the resulting regularity-rooted encounter network plays an important role in various urban environment dynamics like epidemic spreading. Given that most contact network-based spreading models still focus on network topology \cite{Rohani2010,Pastro2001} or small-scale contact processes \cite{Stehle2011,Salathe2010,Isella2011,Rocha2011}, identifying this real-world physical contact network is potentially important in studying encounter patterns and diffusion/spreading dynamics in large populations. To model the dynamical evolution of this contact network, we extracted time-aggregated networks of people with ${w_i} \ge 10$. At the top of Fig.~\ref{Fig3}D, we plotted the fraction of those individuals over the population, finding a rapid increase from 0 to about 90\% on Monday, followed by slower growth afterward. We next checked the variation of average number of encounters $\langle k \rangle $ and average number of encountered people $\langle e\rangle $, finding linear increases of both $\langle k \rangle $ and $\langle 3 \rangle $ without saturation, indicating that the people one may encounter differ from day to day, result in weak and random connections compared to social relations (middle of Fig.~\ref{Fig3}D). However, we note that $\langle k \rangle $ increases faster than $\langle e \rangle $, suggesting that ``stronger'' connections with ``familiar strangers'' are formed gradually through random encounters. At the bottom of the figure, we plotted the evolution of average clustering coefficient $c$, finding that the encounter network displays strong small-world property network with characteristic path length $\langle l \rangle \ge l_{rand} $ and $c\gg {c_{rand}}$ (\cite{Watts1998}, $\langle l \rangle=2.95$, $l_{rand}=\ln\left(n\right)/\ln\left(k\right)=2.63$  and diameter $l_{max}=6$; $c=0.19$ and $c_{rand}=\langle e \rangle/n=4.5\times 10^{-4}$ ). Viewed as a whole, the empirical encounter network we illustrate here is a well-connected small-world graph, in which individuals are no longer confined to local encounters in one vehicle, but interact strongly with increasing number of people across the whole city from day to day.

\section{Discussion}

It has been assumed that human behavior and social interaction/contagion were connected for a long time; however, it is difficult to identify the link between them in observational studies \cite{Christakis2013,Aral2009}. Taking advantage of the availability of metropolitan data collection offered by smart cards in Singapore (this exercise could be very useful for other cities around the world), we tie together thinking on individual mobility with collective interaction patterns. Although the specific results are certainly embedded within the social profile and data of our study, questions remain: how can smart card data give such insight on social interactions and how much do these patterns vary from context to context? As a result of various preferences and constraints on individual behavior, spatial-temporal patterns and collective regularity can be found in daily life, such as morning/evening peaks in transportation, degree of crowding in shopping malls and supermarkets at weekends and in restaurants during dining hours and so forth. Transit use is only one of these social activities with a limited time allocation and specific locations. This study scrutinized only one of a whole spectrum of metropolitan patterns, although it is a critical one. We took transit users as our subjects and the definition of encounter in our study is limited to physical proximity or co-presence, which is characterized by individuals occupying the same vehicle simultaneously. Considering vehicle configuration and loading profile (see SI Appendix section I), physical proximity does not necessarily indicate a more intense social contact, such as talking to each other, but does imply diverse interactions, from not noticing each other, to fleeting eye contact and close observation. As a core of social psychology, social contagion deals with thoughts and behaviors of others by innovation adoption, rumor spreading and decision making. Although the similarity between social contagion and epidemic spreading was recognized a long time ago \cite{Goffman1966}, in the context of physical proximity, social contagion depends more on familiarity than epidemic spreading. Thus, future questions are raised: how to measure the familiarity in the passive ``familiar strangers'' networks and how to define the threshold of familiarity on social diffusion processes. Nevertheless, we know that social communities emerge from the increasing familiarity of individuals into collective forms: from unintentional and passive interactions to intentional and active communication, from mere physical proximity to mutual trust \cite{Grannis2009}. We have shown the existence of city-wide structures of co-presence in Singapore; the extent to which the bus users are aware of these structures has yet to be determined.

Unlike other social networks, where people interact within a circle of friends and acquaintances, we show an often-ignored type of social link: weak, passive and indirectly enabled by daily encounters. As a result of deep-rooted individual behavior patterns, our results also present the collective regularity of people with their recurring encounters as evidence, explaining the ``familiar strangers'' phenomenon in daily life. To our knowledge, this is the first empirical study on the structuring of physical encounters on a metropolitan scale. With the rapid development of information and communication technologies, richer data will be generated throughout our daily life \cite{Vespignani2009}. Although the role of such data is limited by an inherent trade-off: it does not portray everything in detail. However, the emergence of such data provides us with considerable opportunities to enhance our understanding of the social science. Our work should serve as a base to better understand collective human behaviors, dynamical evolution of social networks \cite{Nowak2006,Palla2007}, and especially the impact of collective regularity on various diffusion/spreading processes \cite{Read2008,Meyers2003,Christakis2007}.

\section{Acknowledgments}
We thank Singapore's Land Transport Authority for providing the smart card data, Peng Gong and Song Liang for discussion and comments on the manuscript, Karen Ettlin for copyediting the paper, and Margaret Grieco for giving us important comments on the text. This study was supported by National Research Foundation of Singapore, the funding authority of the Future Cities Laboratory.

\end{document}